\documentclass[format=acmlarge, review=false, screen=true]{acmart}
\pdfoutput=1
\usepackage{booktabs} 
\usepackage[utf8]{inputenc}
\usepackage[ruled]{algorithm2e} 
\usepackage{enumitem}
\usepackage{amsmath}
\usepackage{graphicx}
\usepackage{capt-of}
\usepackage{color}
\usepackage{soul}          
\usepackage[draft,inline,nomargin]{fixme}
\usepackage{multirow}
\usepackage{setspace}
\usepackage{epstopdf}
\usepackage{booktabs}
\usepackage{array}
\usepackage{url}
\input{glyphtounicode}
\usepackage[T1]{fontenc}
\usepackage{subcaption}
\usepackage{tabularx}
\usepackage{ltxtable}
\usepackage{graphicx}
\usepackage{booktabs}
\usepackage{multirow}
\usepackage[normalem]{ulem}
\usepackage{hyperref}
\hypersetup{colorlinks,allcolors=black}
\useunder{\uline}{\ul}{}

\SetAlFnt{\small}
\SetAlCapFnt{\small}
\SetAlCapNameFnt{\small}
\SetAlCapHSkip{0pt}
\IncMargin{-\parindent}

\newcolumntype{L}[1]{>{\raggedright\let\newline\\\arraybackslash\hspace{0pt}}m{#1}}
\newcolumntype{C}[1]{>{\centering\let\newline\\\arraybackslash\hspace{0pt}}m{#1}}
\newcolumntype{R}[1]{>{\raggedleft\let\newline\\\arraybackslash\hspace{0pt}}m{#1}}

\AtBeginDocument{%
  \providecommand\BibTeX{{%
    \normalfont B\kern-0.5em{\scshape i\kern-0.25em b}\kern-0.8em\TeX}}}

\acmVolume{} 
\acmNumber{} 

\setcopyright{none}
\renewcommand\footnotetextcopyrightpermission[1]{}

\begin{document}
%
\title{Using Mobile Data and Deep Models to Assess Auditory Verbal Hallucinations}

\author{Shayan Mirjafari} 
\affiliation{
\institution{Dartmouth College}
\city{Hanover}
\state{NH}
\postcode{03755}
\country{USA}
}
\email{shayan@cs.dartmouth.edu}

\author{Subigya Nepal}
\affiliation{\institution{Dartmouth College}
\city{Hanover}
\state{NH}
\postcode{03755}
\country{USA}
}
\author{Weichen Wang}
\affiliation{%
  \institution{Dartmouth College}
  \city{Hanover}
  \state{NH}
  \postcode{03755}
  \country{USA}
}
\author{Andrew T. Campbell}
\affiliation{\institution{Dartmouth College}
\city{Hanover}
\state{NH}
\postcode{03755}
\country{USA}
}

%

%
\begin{abstract}
Hallucination is an apparent perception in the absence of real external sensory stimuli. An auditory hallucination is a perception of hearing sounds that are not real. A common form of auditory hallucination is hearing voices in the absence of any speakers which is known as Auditory Verbal Hallucination (AVH). AVH is fragments of the mind's creation that mostly occur in people diagnosed with mental illnesses such as bipolar disorder and schizophrenia.
Assessing the valence of hallucinated voices (i.e., how negative or positive voices are) can help measure the severity of a mental illness. We study N=435 individuals, who experience hearing voices, to assess auditory verbal hallucination. Participants report the valence of voices they hear four times a day for a month through ecological momentary assessments with questions that have four answering scales from ``not at all'' to ``extremely''. We collect these self-reports as the valence supervision of AVH events via a mobile application. Using the application, participants also record audio diaries to describe the content of hallucinated voices verbally. In addition, we passively collect mobile sensing data as contextual signals. We then experiment with how predictive these linguistic and contextual cues from the audio diary and mobile sensing data are of an auditory verbal hallucination event. Finally, using transfer learning and data fusion techniques, we train a neural net model that predicts the valance of AVH with a performance of 54\% top-1 and 72\% top-2 F1 score.
\end{abstract}

%
%

%

%

%
\maketitle

\section{Introduction}
\label{sec:intro}

Auditory Verbal Hallucination (AVH) is a perceptual experience of hearing voices that do not exist in the physical world. This form of hallucination may lead to significant distress and impairment~\cite{Mallikarjun2018}. According to epidemiological studies~\cite{Tien1991,Johns2004,deLeedeSmith2013}, AVH occurs in the range of 5-28\% of the general population. AVH is prevalent in people diagnosed with schizophrenia-spectrum disorders, but also occurs in the context of a range of other mental health conditions such as depression and bipolar disorder. However, individuals who are considered mentally healthy (not diagnosed with any sort of mental illnesses) may also hear voices. Up to 75\% of those who report AVH experiences do not meet diagnostic criteria for a psychotic disorder~\cite{2014, deLeedeSmith2013}. AVH is considered to be part of a continuum of psychotic experiences ranging from ``normal'' to ``pathological''~\cite{Johns2001}. There are individuals who experience AVH but may not need a clinical care, while some other individuals suffer from severe and debilitating psychosis and require special care. Identifying the individuals who need a special care needs to be studied further~\cite{goulding2013prodrome, Johns2014}. Recent AVH studies conducted under the Research Domain Criteria (RDoC)~\cite{ford2016studying, iacono2016achieving} framework are primarily focused on neurobiological, physiological and neurocognitive aspects of people with schizophrenia-spectrum disorders~\cite{heckers2009neurobiology, tamminga2000biology, Medalia2008}.

In recent years, mobile sensing is shown to be a good proxy of human behavior, useful for detecting early signs of mental illnesses such as anxiety~\cite{huang2016assessing,saeb2017mobile, info:doi/10.2196/10101, boukhechba2017monitoring}, depression~\cite{Zakaria2019, wang2014studentlife,wang_tracking_2018, mehrotra2016towards, saeb2016relationship}, bipolar disorder~\cite{grunerbl2014smartphone, abdullah2016automatic} and schizophrenia~\cite{ben2017crosscheck,wang2016crosscheck,wang2017predicting}. The increasing number of sensors embedded into mobile devices facilitates us to passively and actively monitor a wide range of behavioral patterns from contextual (e.g., location from GPS, audible sounds around from microphone) to physiological (e.g., heart rate from photoplethysmogram [PPG]) without any user engagement. In addition to sensing, mobile phones are also used to support applications that prompt Ecological Momentary Assessment (EMA)~\cite{moskowitz2006ecological}--a self-report paradigm that involves active and overt data capture. In EMA studies, participants are intervened by mobile applications to complete multiple questionnaires in real-time and in their own environment. 

Linguistic cues in associations with mental-related illnesses are widely studied. ``Subtle features in people’s everyday language may harbor the signs of future mental illness'', researchers state in~\cite{Rezaii2019}. They show the linguistic marker of semantic density and the latent semantic content of an individual’s speech are potentially predictive of psychosis. One of the core features of psychosis is auditory hallucination~\cite{vahia2013diagnostic, bauer2011culture, david1999auditory}. Full auditory hallucination, a positive symptom of psychosis, normally appears relatively late in the course of developing psychosis~\cite{haefner2006early}. However, given the advances in machine learning methods especially natural language processing (NLP), it may be possible to detect the early signs of auditory hallucination as an increased tendency to implicitly talk about voices and sounds~\cite{Rezaii2019}. Authors in~\cite{ratana2019comprehensive} review latest NLP techniques that have been used to characterize various language phenotypes in psychosis, mostly focusing on schizophrenia spectrum disorders (SSD). Speech disturbance is a hallmark of SSD. Tang~\cite{tang2021natural} et al. utilize Bidirectional Encoder Representations from Transformers (BERT)~\cite{devlin2018bert} to explore methods for characterizing speech changes in SSD. Their exploratory results suggest that natural language measures may yield clinically relevant and informative biomarkers of SSD. 

In this study, we focus on assessing the valence of Auditory Verbal Hallucination (i.e., how negative or positive hallucinated voices are) using mobile data. Authors in~\cite{deLeedeSmith2013, Laloyaux2019} state that the higher negative valence of the voices demonstrates the higher possibility of severe psychosis, which requires clinical treatments. We believe our work shows that considering intervention mobile systems is helpful and complementary to further studies on auditory verbal hallucination treatments. To the best of our knowledge, our work as a proof of concept represents the first time that mobile data is used to assess Auditory Verbal Hallucination. 

We study N=435 individuals who experience hearing voices. Participants report the valence of voices they hear four times a day for a month through ecological momentary assessments prompted by a mobile application. We collect this information as the valence supervision. Using the application, participants also record audio diaries to verbally describe the content of hallucinated voices. In addition, we passively collect mobile sensing data as contextual signals. We then experiment with how predictive these linguistic and contextual cues from the audio diary and mobile sensing data are of an auditory verbal hallucination event. We predict the valence of hallucinated voices (scaled from ``Not at all'' to ``Extremely/A Lot'') using linguistic and contextual cues from audio diary and passive mobile sensing data. Specifically, the contributions of this paper are as follows:

\begin{itemize}
    \item We generate auditory features from diary voices using the pre-trained VGGish~\cite{45857, 45611} model. We also transcribe the audio to text using the speech-to-text technique~\cite{khilari2015review}. We then use the pre-trained BERT~\cite{devlin2018bert} model to extract word-to-vector~\cite{mikolov2013efficient} embeddings from the text transcripts as language-related signals.
    \item We propose a new approach that uses VGGish to transform the passive sensing data to a higher-dimensional space. To evaluate how our VGGish-based approach performs, we generate another feature set of mobile sensing data using ROCKET~\cite{dempster2019rocket}. Later, in Section~\ref{sec:results}, we show that the prediction model trained on the features transformed by VGGish outperforms the model trained on the features generated by ROCKET.
    \item We finally report on how predictive the linguistic and contextual cues from the audio diary and mobile sensing data are of an auditory verbal hallucination event. To do so, we train various models based on different data streams as there is multimodality in our feature set (sensing, audio, transcripts). We find that a neural net model using transfer learning and data fusion techniques achieves the highest prediction performance of 54\% top-1 and 72\% top-2 F1 score.
    
\end{itemize}

The structure for the rest of the paper is as follows. In Section~\ref{sec:overview}, we illustrate our study design and data collection in detail. We explain the methodology of modeling the problem, how we generate features for various data streams and the dataset creation in Section~\ref{sec:method}. The architecture of the prediction model and the training process are described in Section~\ref{sec:prediction}. The experimental results of the model's performance when predicting the valence of hallucinated voices are reported in Section~\ref{sec:results}. We discuss the implication of our methodology and describe the prior related work in Section~\ref{sec:discussion} and Section~\ref{sec:related_work}, respectively. Finally, we conclude the paper in Section~\ref{sec:conclusion}.

\section{Study}
\label{sec:overview}

\subsection{Enrollment}
\label{subsec:design}

Participants in the study are enrolled through a website with the help of online advertisement. We advertise the study using Google Adwords which is a contextual advertisement service provided by Google. The advertisements would then appear on websites and blogs that signed up for the ad program and also on Google's search results page, targeted towards US traffic only. Since the ads are contextual, they are likely shown to those people who search for terms that are relevant to auditory hallucination. Please refer to~\cite{buck2021expanding} for details about evaluations of a web-based approach for remote recruitment of people who hear voices.

Would-be participants must pass an initial screening on the website in order to be eligible to participate in the study. The initial screening filters people with auditory verbal hallucination based on clinical criteria put forth by our psychiatry colleagues. Other than the clinical criteria, they must also live in the United States and have an Android phone. Eligible participants receive a download link of our Android application through SMS. After installing the application, participants receive another SMS with a link to automatically login into the application. The link contains a token authenticating the identity of the participant. When the participants click on the link, they will be automatically logged into the application and some information (e.g., model of the phone, Android version, study start date) will be set in our backend server. 

Participants need to stay in the study for a period of 30 days in order to be compensated with \$125. The study is conducted in accordance with the Institutional Review Board (IRB), an institution which protects the rights and welfare of human research subjects. We make sure that participants knowingly consent to be part of the study. During enrollment, participants have the opportunity to ask questions and are given supportive materials, including visuals outlining all sensors and information collected by the app. Prior to obtaining consent, we also test participants' knowledge of the type of data collected, whether the study was confidential, not anonymous, and ways in which we protect their data. Participants are administered a competency screener to verify that they understand the details and are able to provide informed consent. They are not allowed to participate if they could not pass this test. 

\begin{table}[ht]
\begin{minipage}[b]{0.4\linewidth}
\centering
\caption{Diagnoses reported}
\label{tab:diagnosis}
\begin{tabular}{l|l}
\hline
     Diagnosis & Count \\
 \hline
Alzheimer’s or Parkinson’s Disease & 1 \\
Bipolar disorder & 185 \\
Depression & 275  \\
Migraines & 78 \\
Posttraumatic stress disorder (PTSD) & 179   \\
Schizoaffective disorder & 122  \\
Schizophrenia & 126  \\
Seizures & 26  \\
Substance use & 138 \\
Traumatic brain injury (TBI) & 31 \\ 
Other & 82 \\
No answer & 12 \\
\hline
\end{tabular}
\end{minipage}
\hspace{0.5cm}
\begin{minipage}[b]{0.4\linewidth}
\caption{Race distribution}
\label{tab:race}
\begin{tabular}{l | l}
\hline
    Race & Count \\
\hline
American Indian or Alaskan Native & 6 \\
Asian & 7  \\
Black or African American & 100 \\
Pacific Islander & 1  \\
White & 259  \\
More than one race & 52  \\
No answer & 10  \\
\hline
\end{tabular}
\end{minipage}
\end{table}

The number of participants studied in this paper is 435. Majority of them i.e., 232 are women, 189 are men and the remaining 14 report belonging to other genders. With respect to age, 27 participants are under the age of 25. 307 participants are within the age of 25 to 50 and 101 are above 50. Participants are also asked to report any mental illnesses that they might have been diagnosed with. This information helps us better understand the population under study. We report the diagnoses in Table~\ref{tab:diagnosis}. As shown in the table, the total count of diagnoses is higher than the total number of participants, meaning that participants are diagnosed with multiple illnesses. Overall, we find that the majority of our participants are diagnosed with depression. Table~\ref{tab:race} shows the distribution for the participants' race.

\subsection{Ecological Momentary Assessment (EMA)}
\label{subsec:ema}

Participants report their experience of Auditory Verbal Hallucination (AVH) events on a daily basis through EMAs. The EMA query pops up four times a day on participant's phone at varied time-points. In addition to the four time-points, participants could also open the mobile app manually at any other time and report if they experience an auditory verbal hallucination. For the automated EMA notification, we randomly select four time periods within each of these time intervals: 9 to 11 am, 12 to 2 pm, 3 to 5 pm and 6 to 8 pm to send the questionnaire to the participant's phone each day.

\begin{figure}[t]
\centering
\begin{subfigure}[t]{0.4\textwidth}
   \includegraphics[width=\linewidth]{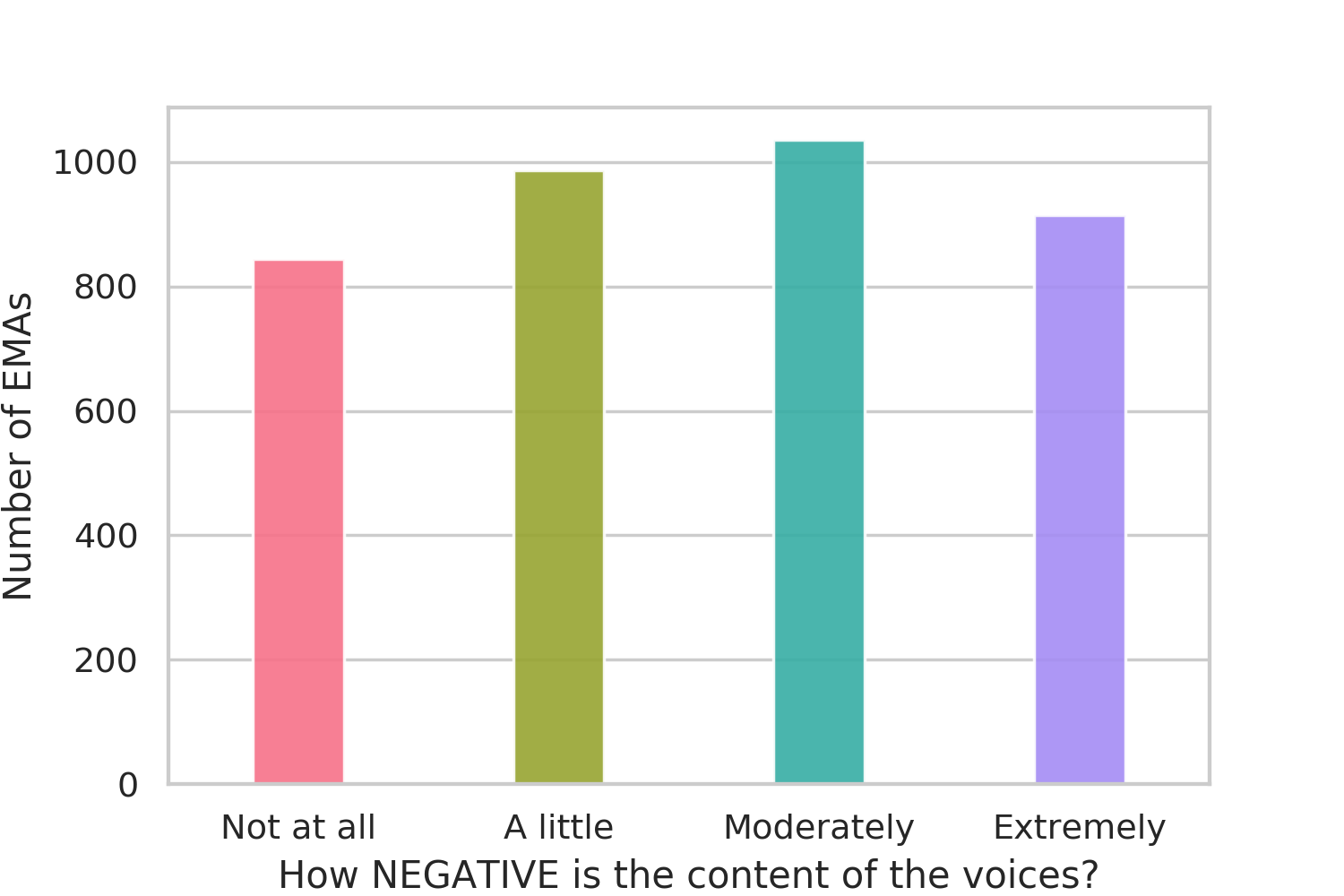}
\end{subfigure}
\hfil
\begin{subfigure}[t]{0.4\textwidth}
   \includegraphics[width=\linewidth]{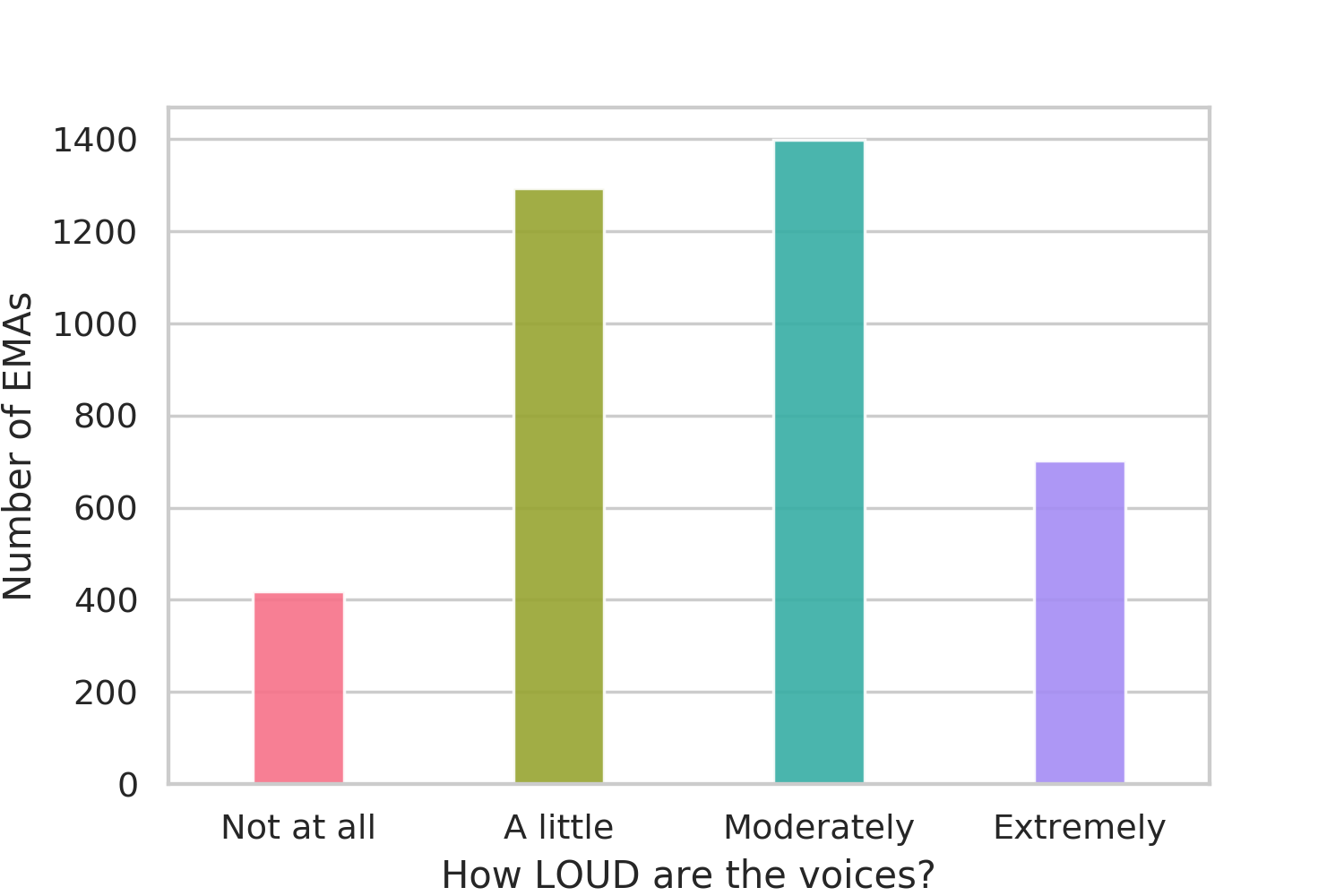}
\end{subfigure}
\begin{subfigure}[t]{0.4\textwidth}
   \includegraphics[width=\linewidth]{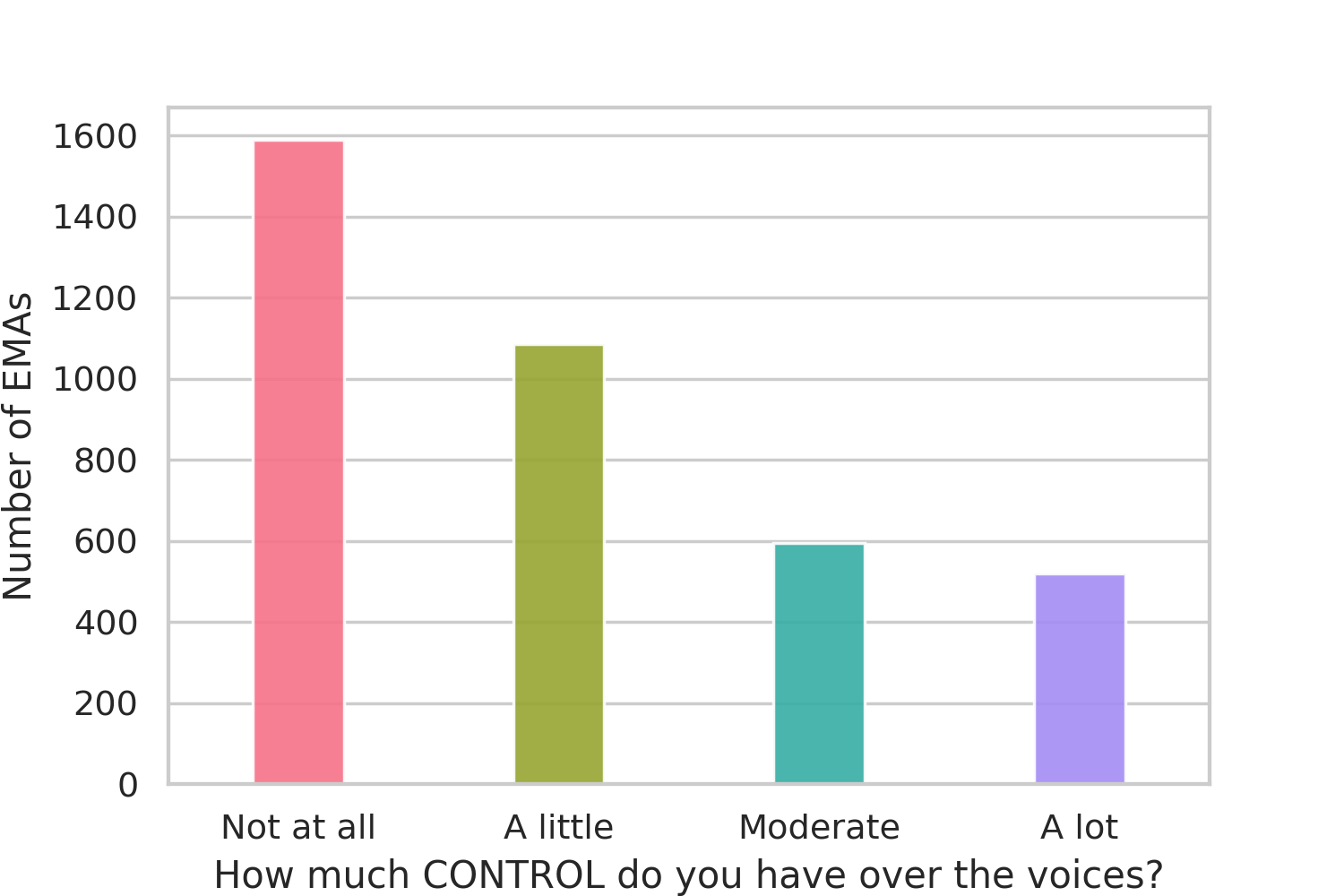}
\end{subfigure}
\hfil
\begin{subfigure}[t]{0.4\textwidth}
   \includegraphics[width=\linewidth]{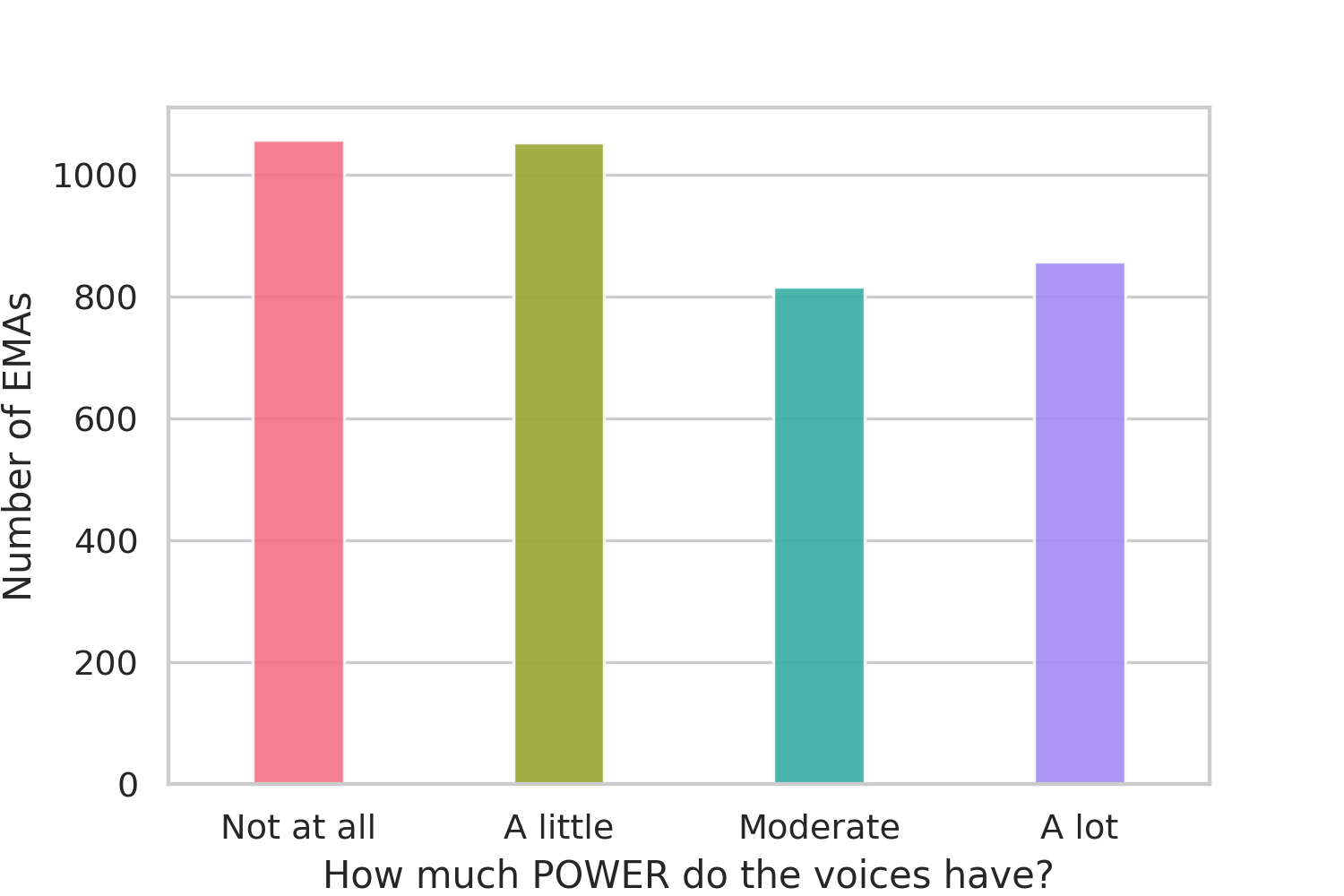}
\end{subfigure}
\caption{Distributions of answers to each of the four EMA questions.}
\label{fig:gt_dist}
\end{figure}

When the participants are prompted with the EMA, we first ask -- \textit{Are you experiencing hearing voices right now?}. If the answer to the question is yes, then the follow-up questions regarding the valence of voices are shown. However, if the answer to the initial question is no, the EMA popup window is instantly closed and there are no further questions.
The follow-up questions are:
\begin{enumerate}
    \item How NEGATIVE is the content of the voices? 
    \item How LOUD are the voices?
    \item How much CONTROL do you have over the voices?
    \item How much POWER do the voices have?
\end{enumerate}
The answer scales for the first two questions are ``Not at all'', ``A little'', ``Moderately'' and ``Extremely'' and for the last two questions are ``Not at all'', ``A little'', ``Moderate'' and ``A lot''. We consider the answers to these four questions as an indicator for the valence of the voices. For example, a group of answers such as extremely negative content, extremely loud, a little bit of control and a lot of power demonstrates the higher possibility for the negative valence of the voices. In the rest of the paper, we refer to the first question as \textit{negativeness}, to the second question as \textit{loudness}, to the third one as \textit{control} and to the last question as \textit{power}, respectively. Figure~\ref{fig:gt_dist} shows the distributions of answers to the EMA questions.

\subsection{Mobile Data}
\label{subsec:mobile}
With mobile phone, we primarily collect two forms of data: active and passive. Active data refers to the audio diary that the participants record detailing their AVH experience. Passive data refers to data that our mobile app collects (i.e., sensors data) in the background.  

\subsubsection{Audio Diary}
If the participants report on the EMA that they have had an Auditory Verbal Hallucination (AVH) experience, we ask them to optionally record an audio diary detailing the content of the AVH. Participants can record an audio diary with a duration of up to three minutes explaining their experience of hallucinated voices and the content of voices. Audio diaries are stored on the phone and uploaded to our secure backend server. 
In the following, we show a quote which is recorded by one of the participants as an example for the diary: ``came outside to let our big dog off of his run and to let him have some freedom. my spouse was across the field for me and I heard a person's voice say can you see it? can you see it? you can see that! and there's no person here. I thought it might be the person chasing cows, but it's not. there's nobody else here, just my spouse, myself and my two children. so that's what it is''. This quote is transcribed from the audio using the Google speech-to-text model. We transcribe all the recorded audio diaries to have a textual asset in our dataset too.

\subsubsection{Passive Sensing}
We collect mobile sensing data through our Android app, which is based on StudentLife~\cite{wang2014studentlife}. The app runs on the background and tracks participant's locations and phone usage. In addition, it also captures proxies for conversations. Table~\ref{tab:sens_feats} expands on the features generated from the mobile sensing data. The app does not record any raw audio information, but there is a speech classifier built into it that checks for the presence of speech in the audio received through the microphone. Using the detected speech, it generates three features: audio amplitude, conversation duration and number of conversations that the participant has been around or has taken part in. Note that we cannot say for certain if it is the participant who is involved in the conversation, but we can infer that they are around some conversation.

\begin{table}[h]
\centering
\small
\caption{Features generated from the mobile sensing data}
\label{tab:sens_feats}
\begin{tabular}{L{3cm}|L{3.5cm}}
\toprule
Sensing Type & Features                                                                                       \\ \hline
\multirow{2}{*}{GPS-based Location}      
             & Number of places visited,                                                                     \\
             & Distance travelled                                                                             \\ \hline
\multirow{2}{*}{Phone Usage}  
             & Unlock duration,                                                                                \\
             & Number of unlocks                                                                               \\ \hline
\multirow{3}{*}{Microphone}   
             & Audio amplitude,                                                                               \\
             & Conversation duration,                                                                          \\
             & Number of conversation                                                                         \\ 
\bottomrule
\end{tabular}
\end{table}

With phone usage information, we generate features such as number of phone locks, unlocks, and the duration of phone usage. We use the GPS data collected through the phone to compute significant locations. In order to calculate these locations visited by the participants, we use the DBSCAN~\cite{ester1996density} clustering algorithm. We consider the central coordinate of each cluster as a significant location if the participant has spent at least 30 minutes at that location. We then calculate the number of such locations and the distance traveled between them. We use these sensing features as contextual data.

\section{Methodology}
\label{sec:method}
In this section, we discuss our modeling methodology i.e., the supervision of the problem and how we generate features from different types of data (audio diary and passive sensing data) to create a dataset for the prediction task. 

\subsection{Supervision}
\label{subsec:ground_truth}
As discussed in Section~\ref{sec:overview}, participants are asked to respond to EMAs four times a day to assess the valence of their AVH experience. Each question has four answer scales that we use as the labels/categories for the training process. In total, there are 3838 self-reports from 435 participants. Figure~\ref{fig:gt_dist} illustrates the distributions of answers to each of the four questions. During the training process, we encode these categories using the one-hot encoding technique. We encode ``Not at all'' as [1, 0 , 0, 0], ``A little'' as [0, 1, 0, 0], ``Moderately''/``Moderate'' as [0, 0, 1, 0] and ``Extremely''/``A lot'' as [0, 0, 0, 1]. Later, in Section~\ref{sec:prediction}, we predict these labels for each of the four EMA questions separately.


\subsection{Audio Diary Features}
\label{subsec:diary}
As discussed earlier in Section~\ref{sec:overview}, after answering the EMA questions, participants record an audio diary explaining the content of the voices they hear. We use transfer learning techniques to generate features from the audio for the prediction.
\subsubsection{VGGish}
\label{subsubsec:vggish}
We generate auditory signals from the pre-trained VGGish model~\cite{45857, 45611} by Google which is a convolutional neural network, and is trained on 8M YouTube videos for the sound classification task. VGGish extracts a high-level 128-D embedding of an audio input which can be fed as a feature input to a shallower model for any sound-related classification tasks as the VGGish embedding is more semantically compact than raw audio features~\cite{45857, 45611}. We extract the embedding of the audio diary from VGGish and use that as one of the features for the valence classification task in our problem. Note, before feeding the audio diary into the VGGish model, we first zero-mean the audio frequency data per audio file. As a requirement for the VGGish model, the mel-frequency cepstrum (MFC)~\cite{Xu2004HMMBasedAK} must be first generated from audio formats and then the logit function must be applied to MFC. VGGish features are extracted from non-overlapping audio log-mel patches of 0.96 seconds, where each patch covers 64 mel bands and 96 frames of 10 milliseconds~\cite{45857, 45611}. The output of VGGish is a 128-D embedding vector for each audio patch. We take an average of all the patches' embeddings to generate a 128-D embedding representation for the entire audio file. In Section~\ref{sec:prediction}, we use these 128-D audio embeddings as input features to our prediction model.
 
\subsubsection{BERT}
\label{subsubsec:bert}
To generate textual features, we transcribe the audio-diary using speech-to-text techniques~\cite{khilari2015review}. We then use the pre-trained BERT (Bidirectional Encoder Representations from Transformers)~\cite{devlin2018bert} model to extract contextual word-to-vector embeddings from the transcript sentences. To do so, We first tokenize words in different sentences based on the BERT's vocabulary and feed them into the encoders (transformer layers) of BERT. We then extract a contextualized embedding for each of the input words in a sentence. To generate a representation of an entire transcript (i.e., consists of multiple sentences), we aggregate the encoder outputs after multi-head attention layers of BERT. In the BERT architecture we used, there are 12 separate 768-D encoder outputs for every single token of the input~\cite{devlin2018bert}. With regard to an input text representation, BERT authors have evaluated the model performance using different encoder outputs combinations as input to a further deep model on a named entity recognition task. Based on their reported results, we choose to sum up all the 12 output vectors to get a single 768-D vector representation for an entire sentence in the text. We finally take an average of all sentences' vectors in a transcript to generate a global vector representation for the entire transcript. Later we use this 768-D vector as input features (language-related signals) to our prediction model. 

\subsection{Mobile Sensing Features}
\label{subsec:sensing}
As shown in Table~\ref{tab:sens_feats}, there are seven main sensing features from mobile phone. We compute hourly aggregated features for each sensing feature. For the purpose of the prediction, we consider continuous sensing data within 24 hours prior to an EMA response -- the reason we consider 24 hours of sensing data prior to an EMA is that the questions in the EMA assess the momentary valence of the hallucinated voice by asking participants whether they are experiencing an auditory hallucination \textit{at the moment}, as mentioned in Section~\ref{sec:overview}. Therefore, we assume closer time windows to an auditory hallucination experience conveys more relevant signals about the experience. We generate two different sets of features from the mobile sensing data. 1) We transform the 24-hour mobile sensing time series to banded spectrograms. We then feed these converted signals to the VGGish model and generate a 128-D embedding vector for them. 2) We use an approach called ROCKET (Random Convolutional Kernel Transform)~\cite{dempster2019rocket} which uses random 1-D convolutional kernels~\cite{kiranyaz20191d} to generate higher-dimensional patterns from a time series. We use these two feature sets of mobile sensing in our modeling. \textit{The reason we transform the sensing data to a higher-dimensional space is to make the prediction model have higher degrees of freedom (more power) on the feature set to better estimate the output.} Later, in Section~\ref{sec:results}, we show that these transformations boost the model's prediction performance. We also compare how the prediction model performs when we utilize the transformed features using VGGish from the first approach versus when utilizing the transformed features using ROCKET from the second approach. In the following, we explain the transformations of the sensing features in detail.

\begin{figure}[t]
\centering
\begin{subfigure}[t]{0.3\textwidth}
   \includegraphics[width=\linewidth]{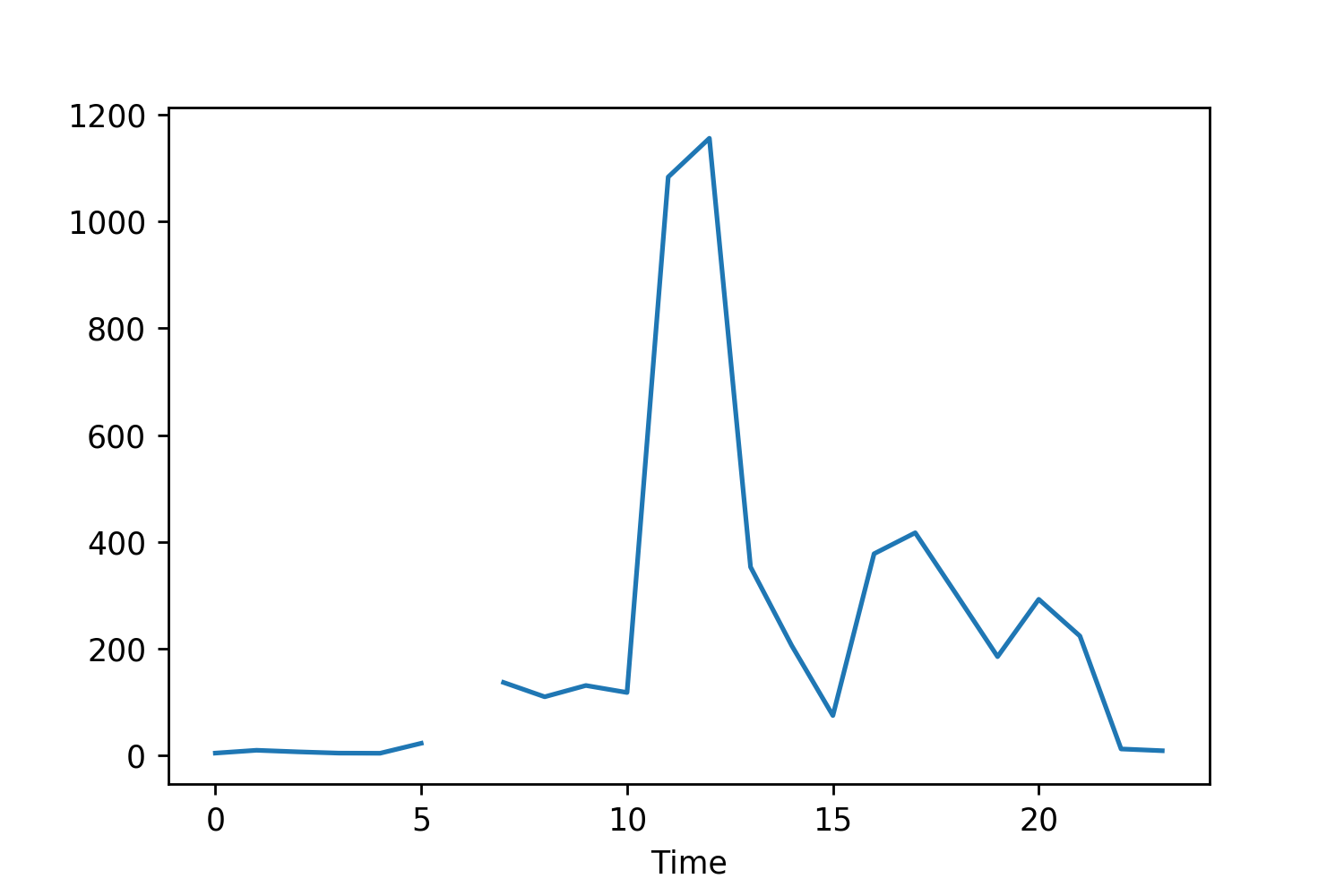}
   \caption{}
\end{subfigure}
\hfil
\begin{subfigure}[t]{0.3\textwidth}
   \includegraphics[width=\linewidth]{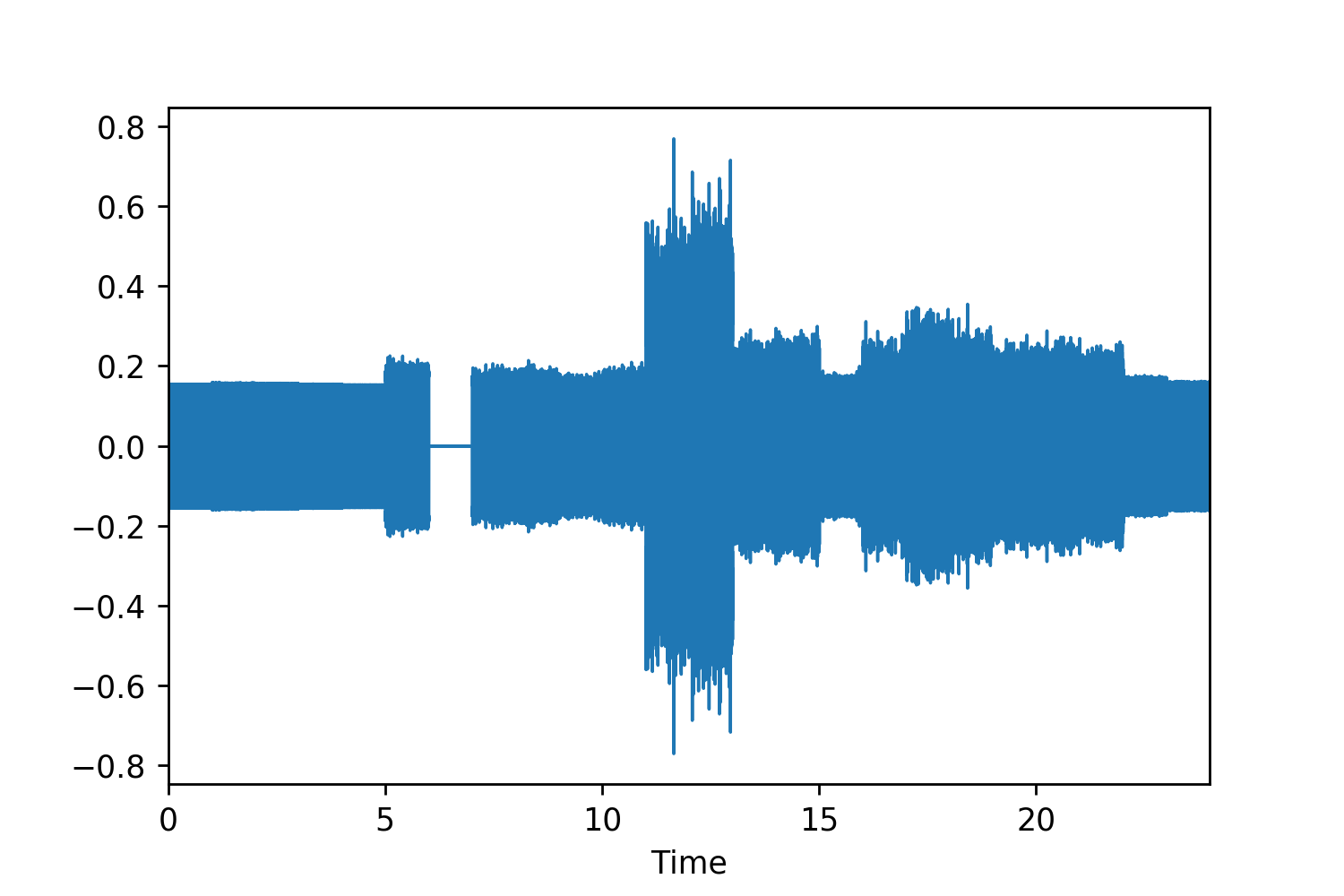}
   \caption{}
\end{subfigure}
\hfil
\begin{subfigure}[t]{0.3\textwidth}
   \includegraphics[width=\linewidth]{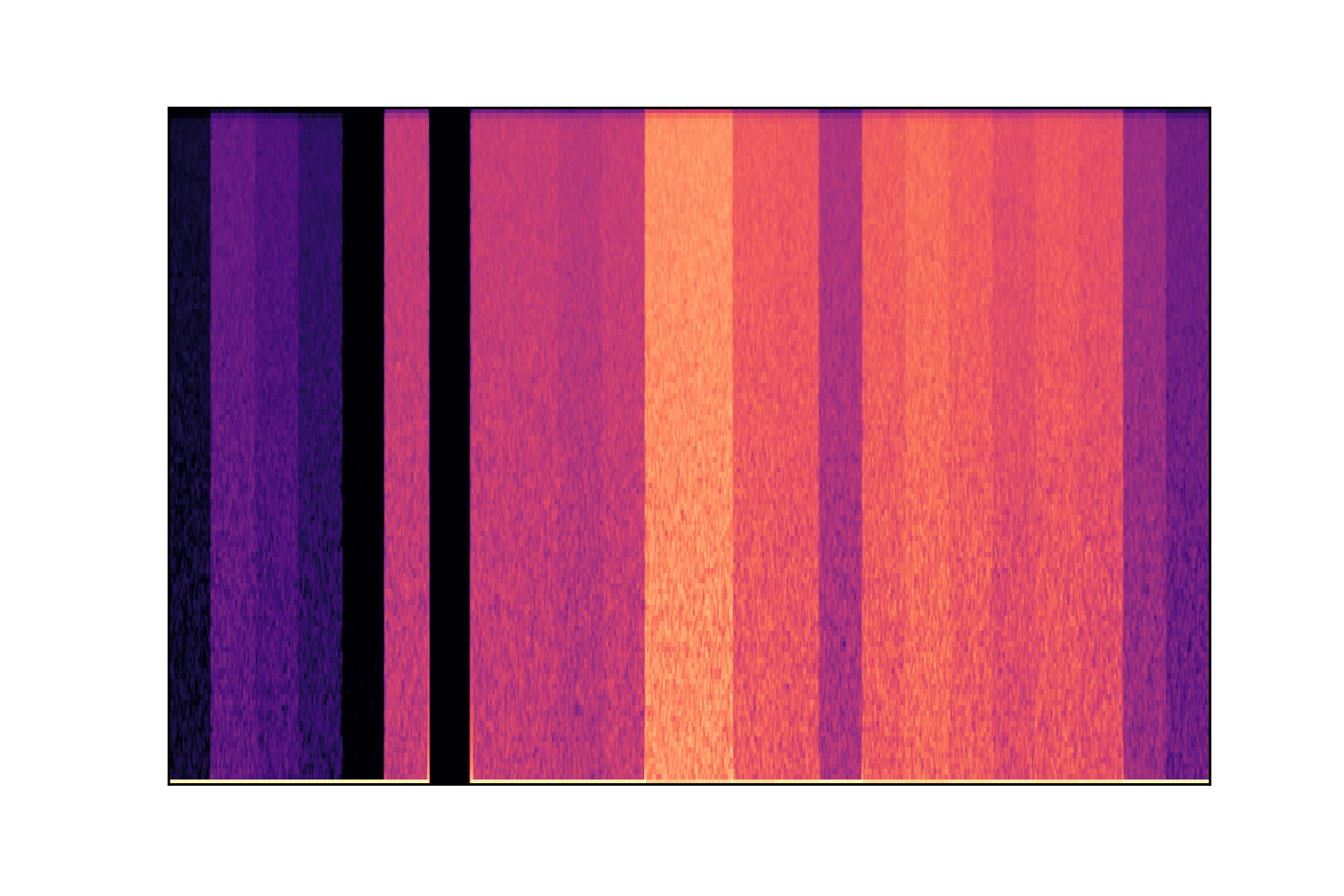}
   \caption{}
\end{subfigure}
\caption{Figure (a) shows an hourly mobile sensing time series for 24 hours prior to an EMA. Figure (b) shows the zero-meaned transformed time series in the time domain. The transformed time series has a length of 24 seconds and a sample rate of 44.1 kHz. Finally, Figure (c) shows the spectrogram of the transformed time series. The VGGish model takes the spectrogram as input and applies 2-D convolution operations to generate a representative higher-dimensional feature set, as discussed in \ref{subsubsec:vggish}. As illustrated, the slices in the spectrogram are representations of the data points in the original time series. The higher the value of the data point in the original time series, the brighter the slice in the spectrogram.}
\label{fig:audio_convert}
\end{figure}

\subsubsection{Time Domain Transformation}
\label{subsubsec:audio_encode}
To generate features from the mobile sensing time series using the VGGish model, we transform the sensing time series to the time domain. As there are 24 hourly data points in the time series prior to an EMA, we generate series in the time domain with a length of 24 seconds and a sampling rate of 44.1 kHz. To do so, we first zero-mean the time series and scale them into the [-1, 1] interval using the min-max scaling. Then, for each data point in the time series, we generate 44100 random data points (as the sampling rate is 44.1 kHz) drawn from a Gaussian distribution with a mean value equal to the time series corresponding point and a small standard deviation around the mean value. The following formula shows the distribution of data points we randomly generate in the time domain. 
\[
    Y \sim \mathcal{N}(\mu=x_{i},\,\sigma^{2}=\epsilon \cdot x_{i})\
\]
$Y$ represents one second of the randomly generated audio and $x_{i}$ represents the value of one single data point in the original time series. The $\epsilon$ is a coefficient that is multiplied by $x_{i}$ to represent the variance of the generated data. The $\epsilon$ is chosen by considering the magnitude of values in the original time series. As the values in our mobile sensing time series are usually small integers (e.g., number of times of unlocking phone per hour), we set $\epsilon$ to be 0.1. As mentioned, the sampling rate is 44.1 kHz. However, it could be a different number than 44.1 kHz. However, because we randomly generate the audio data, a larger hertz increases the possibility that the mean of randomly generated data is closer to the corresponding value in the original time series. Figure~\ref{fig:audio_convert} illustrates how the time series are transformed to the time domain and banded spectrograms. After transforming sensing time series, we feed the spectrograms of them into the pre-trained VGGish model to extract higher-dimensional features. 
As mentioned earlier, the output of VGGish is a 128-D embedding vector. 
Later, in Section~\ref{sec:results}, we show that these embeddings of the transformed sensing time series help boost the performance of the prediction model. 


\subsubsection{ROCKET}
\label{subsubsec:rocket}
The approach used by ROCKET (Random Convolutional Kernel Transform)~\cite{dempster2019rocket} is to generate features using 1-D convolutional kernels~\cite{kiranyaz20191d}. 
However, unlike convolutional neural nets where the convolution kernels are typically learned, ROCKET uses random convolutions. Instead of learning the kernels, they are randomly generated to extract patterns over the time series. Authors show that the approach provides significant improvement over state-of-the-art performance with huge reductions in time and computation power required~\cite{dempster2019rocket} for univariate time series classification. The only hyperparameter for ROCKET is the number of random kernels to generate. ROCKET generates two aggregated features from each feature input -- the maximum value (which is equivalent to global max pooling~\cite{oquab2015object}) and the proportion of positive values per kernel. We use ROCKET to extract these features from the 24-hour mobile sensing time series. The reason we use ROCKET to generate another set of transformed features from the mobile sensing data is to make a comparison on how the prediction model performs in the prediction task using our proposed approach on the VGGish-based transformed features, as described above, versus when using the ROCKET-based transformed features. Therefore, since we generate a 128-D embedding vector from the transformed sensing time series using VGGish, here, we use 64 random kernels for ROCKET to generate exactly 128 features for the time series too. We generate 128 features using ROCKET to have a fair performance comparison with features generated by VGGish. Later, in Section~\ref{sec:results}, we show that the prediction model trained on the feature set generated by the VGGish model outperforms the model trained on the features generated by ROCKET.
\section{Prediction}
\label{sec:prediction}

\begin{figure}[t]
\centering
\begin{subfigure}[t]{.9\textwidth}
   \includegraphics[width=\linewidth]{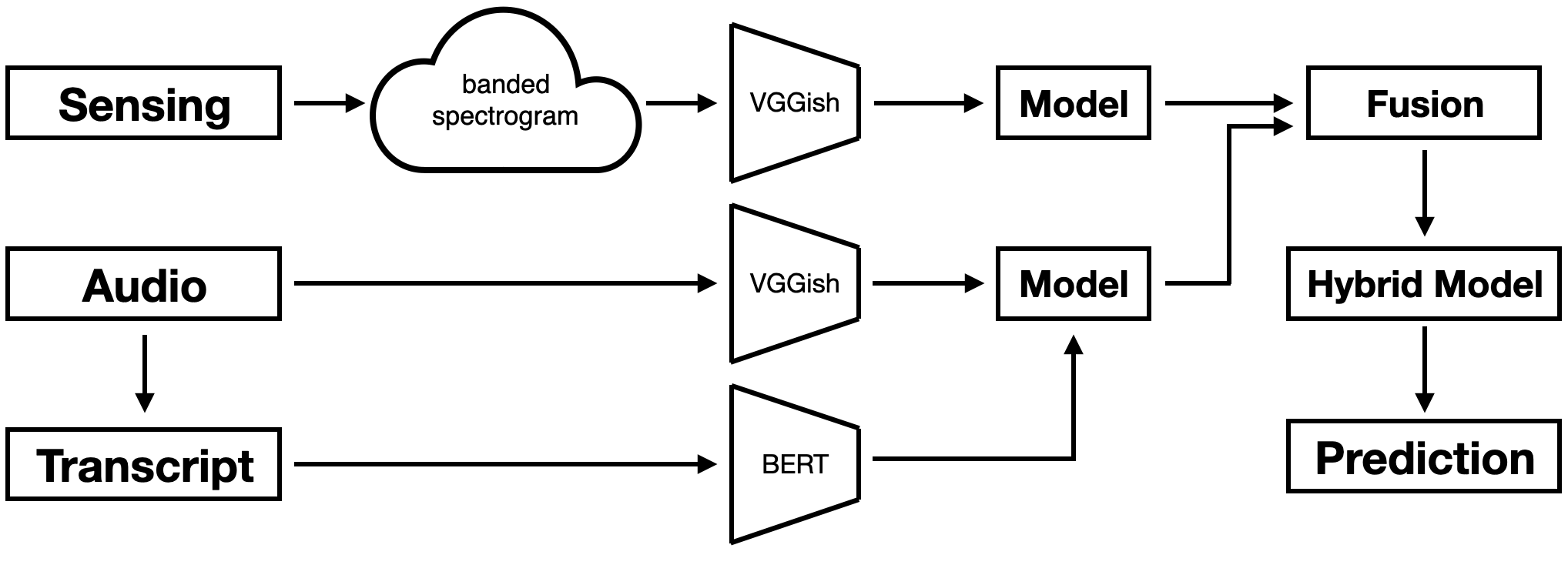}
\end{subfigure}
\caption{Modeling pipeline. We generate auditory and textual (speech-to-text) features from audio diaries using VGGish and BERT, respectively. To improve the prediction performance, we generate higher-dimensional features from sensing time series. We transform the sensing time series to the time domain and then generate features by feeding the banded spectrograms to VGGish -- see Section~\ref{sec:method} for details. Finally, we train fully-connected models using these feature sets for the prediction task. Specifically, using transfer learning and data fusion techniques, we train a hybrid model which obtains the highest prediction performance.}
\label{fig:pipeline}
\end{figure}

After generating the features and creating the dataset, we train different neural networks with various combinations of fully connected layers using the feature sets (auditory, textual and sensing). Figure~\ref{fig:pipeline} shows the overall modeling pipeline. We train three models separately to predict the EMA responses: one model using only auditory and textual features obtained from the audio diaries, another using only mobile sensing transformed features and lastly, a hybrid model that uses a combination of all the feature sets. In Section~\ref{sec:results}, we report prediction performance of all these models and show that the hybrid model outperforms others.
Before training, we split the dataset into training, validation and test sets. To do so, we choose the earliest 60\% of each participant's data for training, then 20\% for validation and finally the remaining 20\% for the test. We keep these sets the same for all the training scenarios to have fair comparisons for the prediction performances. To evaluate, we experiment running models with various architectures and hyperparameters. We finally report the model that performs the best in predictions (i.e., in terms of F1 score) on the held-out test set. The models are to predict the four EMA questions (negativeness, loudness, control and power). In the following, we discuss the architectures of the models for each feature.

\subsection{Auditory-textual features based model}
\label{subsec:pred_audio_text}
As mentioned earlier in \ref{sec:method}, we generate auditory and textual features from the audio diary using VGGish and BERT models, respectively. The size of the auditory feature set generated by VGGish is 128 and the size of the textual feature set generated by BERT is 768. Therefore, the total input feature size is 896. We train a neural model on these features to understand their contribution to the model performance when predicting the valence of the verbal hallucinations. Table~\ref{tab:audio_text_arch} shows the architecture of the fully-connected model. As there are four different categories for the prediction, the output layer has a size of 4 with a sigmoid function which identifies the probability for each label. We use Cross-Entropy for the loss function and Adam~\cite{kingma2014adam} as the optimizer during the training process. To avoid overfitting, we regularize the architecture by using Batch-Normalization~\cite{ioffe2015batch} and Dropout~\cite{hinton2012improving} on the layers. We train the model with a batch-size of 64 for 80 epochs. Later in Section~\ref{sec:results}, we report the model prediction performance.

\begin{table}[h]
\centering
\caption{The architecture of the model trained on auditory-textual features.}
\label{tab:audio_text_arch}
\begin{tabular}{|l|l|l|c|c|}
\hline
layer type & size & activation function & dropout rate & batch normalization \\ \hline
input & 896 &  & 0.6 & \checkmark \\ \hline
hidden & 512 & relu & 0.4 & \checkmark \\ \hline
hidden & 128 & relu & 0.2 & \checkmark \\ \hline
hidden & 32 & relu &  & \checkmark \\ \hline
hidden & 16 & relu &  & \checkmark \\ \hline
hidden & 8 & relu &  & \checkmark \\ \hline
output & 4 & sigmoid &  &  \\ \hline
\end{tabular}
\end{table}

\subsection{Mobile sensing features based model}
\label{subsec:pred_sens}

We generate a 128-D feature set for the transformed sensing data from VGGish and ROCKET, as discussed in Section~\ref{sec:method}. There are seven main mobile sensing streams, as shown in Table~\ref{tab:sens_feats}. Hence, the size of sensing related features is a total of 896 (i.e., 7x128). We train a fully-connected neural network using these features. Table~\ref{tab:sens_arch} shows the architecture of the network. The network uses the cross-entropy loss function and Adam optimizer. We also regularize the model using batch normalization and dropout. The batch-size and epochs for training the network are 42 and 120, respectively.

\begin{table}[h]
\centering
\caption{The architecture of the model trained on sensing related features.}
\label{tab:sens_arch}
\begin{tabular}{|l|l|l|c|c|}
\hline
layer type & size & activation function & dropout rate & batch normalization \\ \hline
input & 896 &  & 0.5 & \checkmark \\ \hline
hidden & 512 & tanh & 0.2 & \checkmark \\ \hline
hidden & 128 & tanh & 0.2 & \checkmark \\ \hline
hidden & 32 & tanh &  & \checkmark \\ \hline
hidden & 16 & tanh &  & \checkmark \\ \hline
output & 4 & sigmoid &  &  \\ \hline
\end{tabular}
\end{table}

\subsection{Fusion based model (hybrid)}
\label{subsec:pred_transfer}

After training the auditory-textual based and sensing based models, we use data fusion techniques to train a hybrid model. We extract and transfer the values given by further layers of the each trained models as input for the hybrid model. We run experiments to train the hybrid model using the output of different layers of the two trained models. We find that the outputs of the layers with the size of 32 provides the hybrid model with the highest prediction performance. Later in section~\ref{sec:results}, we compare the prediction performance of the hybrid model with that of the other individual models. Note that, to evaluate the model performance, we extract these 32 transferred features from each model for the test set by feeding the original test set into the trained models; then, we use these transferred features for the test set to measure the performance of the model. As we extract 32 transferred features from each model, the input size for our hybrid model is 64 (32 from the auditory-textual model, 32 from the sensing model). Table~\ref{tab:transfer_arch} illustrates the architecture of the hybrid model. The output has a size of 4 as that is the number of the categories, followed by a softmax activation function. The optimizer and the loss function of the model are Adam and cross-entropy, respectively. We train the hybrid model with a batch-size of 32 for 50 epochs.

\begin{table}[h]
\centering
\caption{The architecture of the hybrid model trained on transferred features from the auditory-textual and sensing models.}
\label{tab:transfer_arch}
\begin{tabular}{|l|l|l|c|c|}
\hline
layer type & size & activation function & dropout rate & batch normalization \\ \hline
input & 64 &  & 0.5 & \checkmark \\ \hline
hidden & 32 & tanh & 0.3 & \checkmark \\ \hline
hidden & 16 & relu & 0.2 & \checkmark \\ \hline
output & 4 & softmax &  &  \\ \hline
\end{tabular}
\end{table}

\subsection{Overall model}
\label{subsec:pred_all}

We also train a model from scratch on the entire feature set (textual, auditory and sensing) -- we call this overall model. The input feature size of the overall model is 1792. Later in Section~\ref{sec:results}, we compare the prediction performance obtained by this model with other models' performances, especially with the performance of the hybrid model. We show that the hybrid model outperforms the overall model. The architecture of the overall model is shown in Table~\ref{tab:overall_arch}. Same as other models, we choose the cross-entropy loss function and Adam optimizer. The training batch-size is 64 for 150 epochs.

\begin{table}[h]
\centering
\caption{The architecture of the overall model trained on all the textual, auditory and sensing features.}
\label{tab:overall_arch}
\begin{tabular}{|l|l|l|c|c|}
\hline
layer type & size & activation function & dropout rate & batch normalization \\ \hline
input & 1792 &  & 0.6 & \checkmark \\ \hline
hidden & 896 & relu & 0.5 & \checkmark \\ \hline
hidden & 596 & relu & 0.2 & \checkmark \\ \hline
hidden & 128 & relu & 0.3 & \checkmark \\ \hline
hidden & 32 & relu & 0.1 & \checkmark \\ \hline
hidden & 16 & relu &  & \checkmark \\ \hline
hidden & 8 & relu &  & \checkmark \\ \hline
output & 4 & sigmoid &  &  \\ \hline
\end{tabular}
\end{table}
\section{Experimental Analysis}
\label{sec:results}
In this section, we report the models' performances on predicting each of the EMA questions (i.e., negativeness, loudness, control and power). We use F1 score as our evaluation criteria. The models predict four categories for each EMA question (i.e.,``Not at all'', ``A little'', ``Moderate/Moderately'' and ``Extremely/A lot'') using different sets of features (auditory, textual and passive sensing) as discussed in Section~\ref{sec:method}. We experiment with different models based on various data streams to investigate how predictive a specific data stream or a combination of streams is of an AVH event.

Auditory features and textual features are generated by VGGish and BERT, respectively. We also extract higher-dimensional features for mobile sensing data using two approaches discussed in \ref{subsec:sensing}: one, by transforming the sensing time series to banded spectrograms and then feeding into the VGGish model, and another, by using ROCKET which generates features based on 1-D convolutional operations with random kernels. We further compare the prediction results of the sensing model when trained on the VGGish-based transformed features and when trained on the ROCKET-based transformed features.

\begin{figure}[t]
\centering
\begin{subfigure}[t]{0.8\textwidth}
   \includegraphics[width=\linewidth]{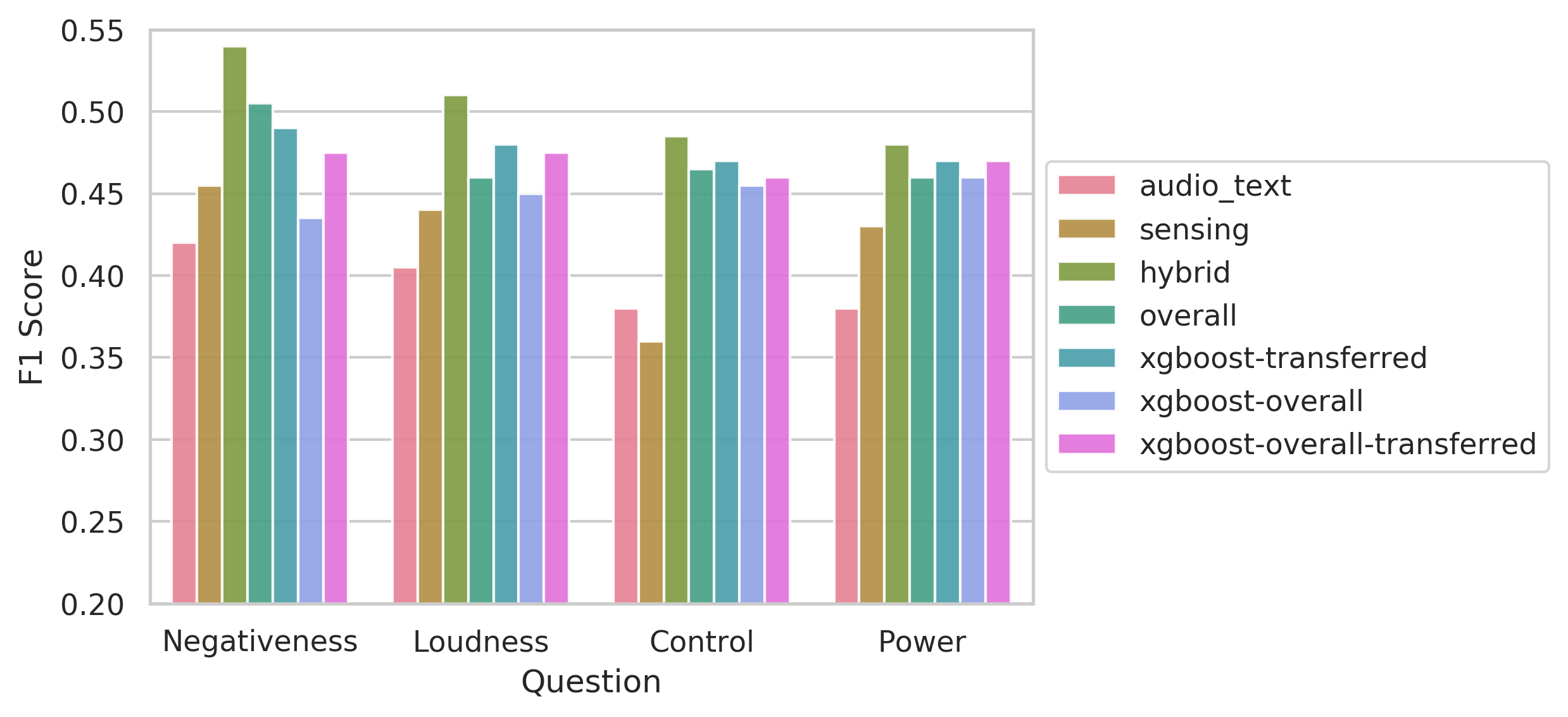}
\end{subfigure}
\caption{The top-1 F1 scores obtained from the trained models when predicting the categories of each EMA question. As shown, the hybrid model outperforms other models.}
\label{fig:top1_vggish}
\end{figure}

\begin{figure}[t]
\centering
\begin{subfigure}[t]{0.8\textwidth}
   \includegraphics[width=\linewidth]{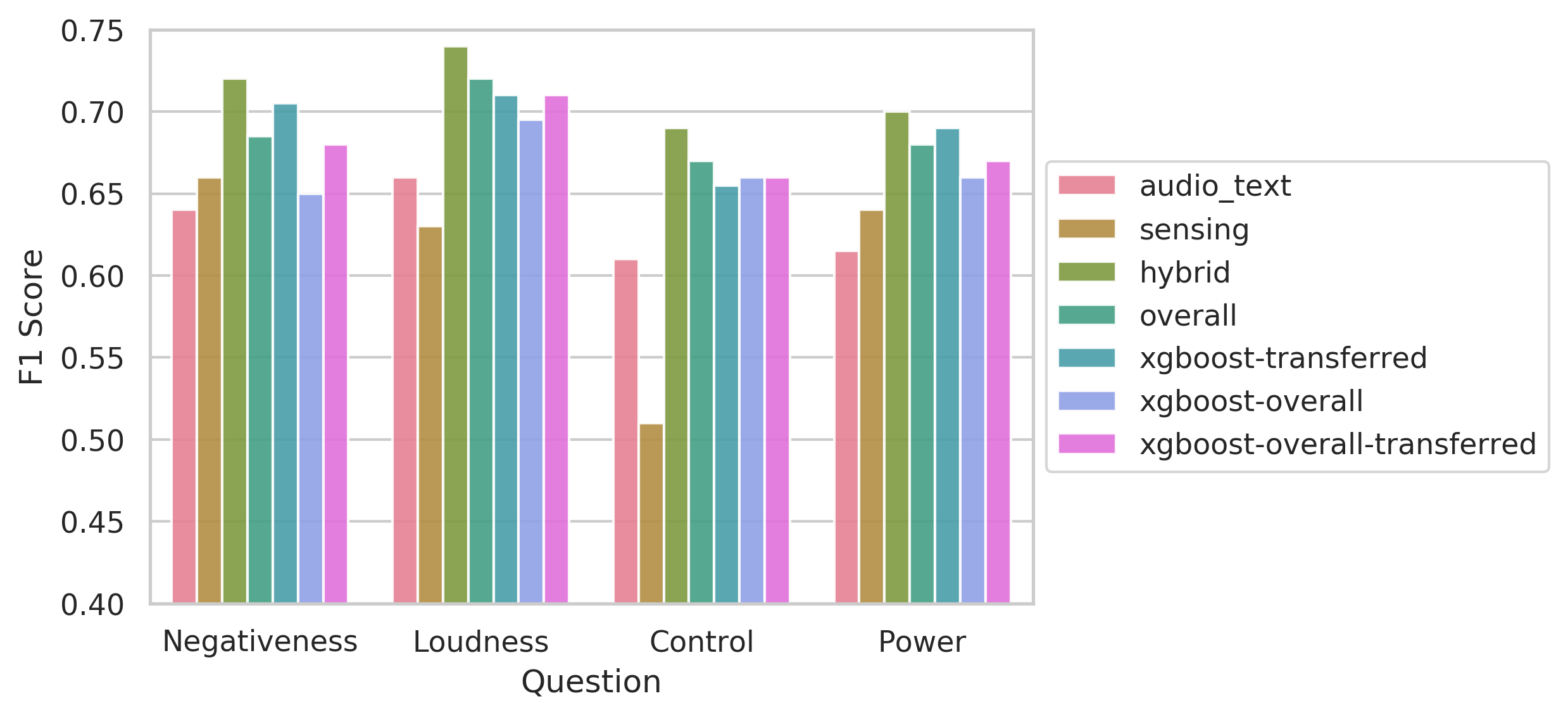}
\end{subfigure}
\caption{The top-2 F1 scores. The top-2 refers to a category predicted by the model with the highest or second-highest probability. As shown, the hybrid model outperforms others.}
\label{fig:top2_vggish}
\end{figure}

We report the results of the prediction performances in terms of top-1 and top-2 F1 scores on the test set for the models discussed in Section~\ref{sec:prediction}). As shown in Figure~\ref{fig:top1_vggish} and \ref{fig:top2_vggish}, in all the training scenarios, the hybrid model outperforms the other models. We also train XGBoost models~\cite{chen2016xgboost} (i.e., decision-tree-based traditional Machine Learning algorithm) using the same sets of features to get performance a baseline for the prediction task using a non-neural model. We train the XGBoost models on three sets of features: a) on all the combined features from the audio, text and mobile sensing (termed xgboost-overall), b) on the transferred features from the trained auditory-textual and sensing model (termed xgboost-transferred), and c) on the  transferred features from the third hidden layer of the overall model (termed xgboost-overall-transferred). As shown, the hybrid model outperforms all the other models, especially the overall model. We use data fusion techniques for the hybrid model as there is multimodality in our feature set. We transfer features from the trained audio-text and sensing models and then combine these features to train the hybrid model. As the transferred features are extracted from the trained models, they are tuned and have learned representations specific to the prediction problem. Therefore, using such the feature set helps the hybrid model perform better. We later, in Section~\ref{sec:discussion}, discuss the reason that we transform the feature sets to higher-dimensional spaces.

The hybrid model obtains a top-1 F1 score of 54\% and a top-2 F1 score of 72\% for the \textit{negativeness} EMA question. For the \textit{loudness} question, the hybrid model obtains a top-1 F1 score of 51\% and a top-2 of 74\%. Similarly, the hybrid model obtains a top-1 F1 score and a top-2 score of 48\% and 68\%, respectively, regarding the \textit{control} question. Lastly, the model performs with a top-1 F1 score of 47\% and a top-2 F1 score of 70\% on the \textit{power} EMA question. Note that, as there are four categories for each question, the chance performance of the top-1 and top-2 scores would be 25\% and 50\%, respectively, if the categories distribution was balanced. However, as illustrated in Figure~\ref{fig:gt_dist}, the distributions are imbalanced. Therefore, we consider a chance model that always predicts the category with the highest occurrence in the test set. Therefore, the top-1 chance performances for \textit{negativeness}, \textit{loudness}, \textit{control} and \textit{power} are 29\%, 38\%, 41\% and 28\%, respectively.

\begin{figure}[t]
\centering
\begin{subfigure}[t]{0.9\textwidth}
   \includegraphics[width=\linewidth]{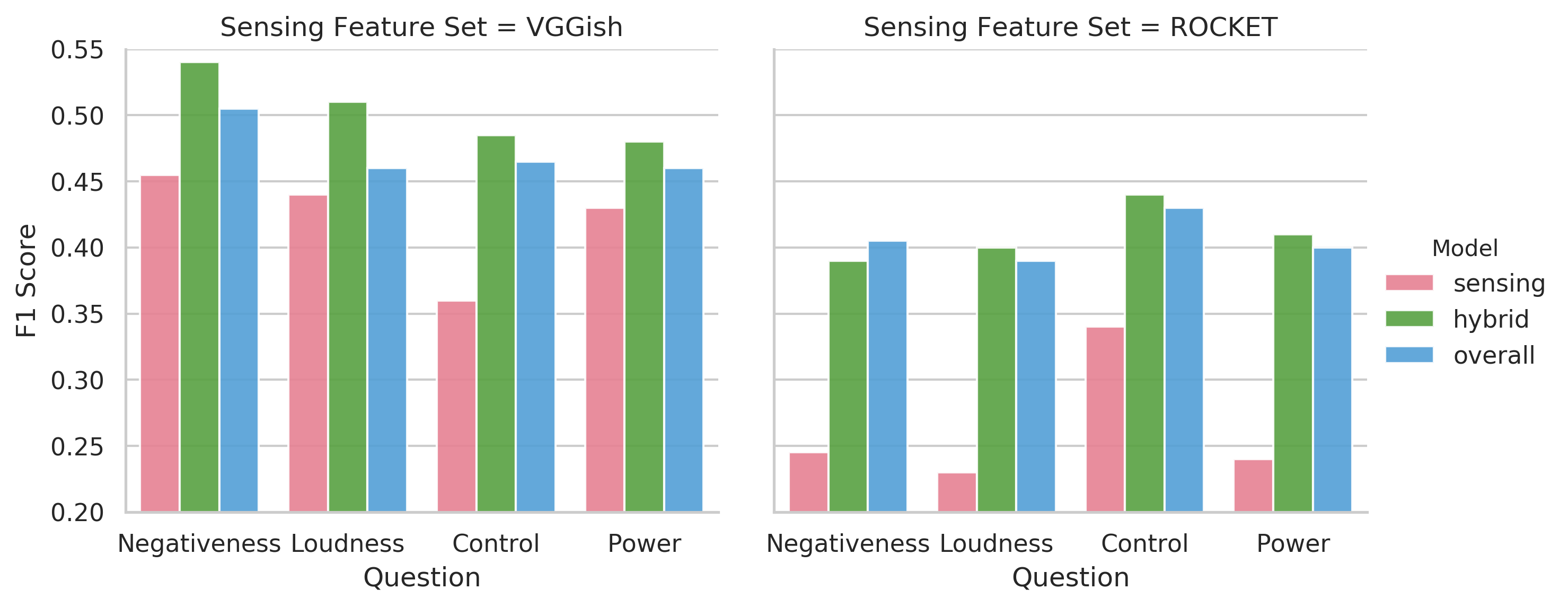}
\end{subfigure}
\caption{The top-1 F1 scores for the models trained using the transformed sensing features. The left figure shows the performances based on the VGGish-based transformed features. The right figure are the performances based on the features generated by ROCKET.}
\label{fig:top1_both}
\end{figure}

In the figures, the sensing features are referring to the VGGish-based transformed features (termed sensing), as the models obtain a better prediction performance on them rather than the ROCKET-based transformed features. As mentioned in Section~\ref{sec:method}, we propose a new approach for transforming the mobile sensing time series to the time domain and banded spectrograms in order to extract features from the pre-trained VGGish model. VGGish has been historically used to extract meaningful features from audio data for different sound recognition tasks. We feed the transformed sensing time series data into the VGGish model and extract features in a higher-dimensional space. We then train the sensing model based on these extracted features. However, to see how much these features contribute to improving the prediction performance, we use ROCKET technique to generate another set of features from the sensing data. ROCKET uses 1-D convolutional operations with random kernels to transform input features. As discussed in Section~\ref{sec:method}, the size of the features extracted from VGGish is 128. Therefore, we generate a feature set with the size of 128 using ROCKET to have a fair comparison between these two approaches. Figure~\ref{fig:top1_both} shows the prediction performance in terms of top-1 F1 score for three different models. As shown, the models i.e., sensing, hybrid and overall trained on the features extracted from VGGish outperform the models trained on the ROCKET-transformed features.

\begin{figure}[t]
\centering
\begin{subfigure}[t]{0.9\textwidth}
   \includegraphics[width=\linewidth]{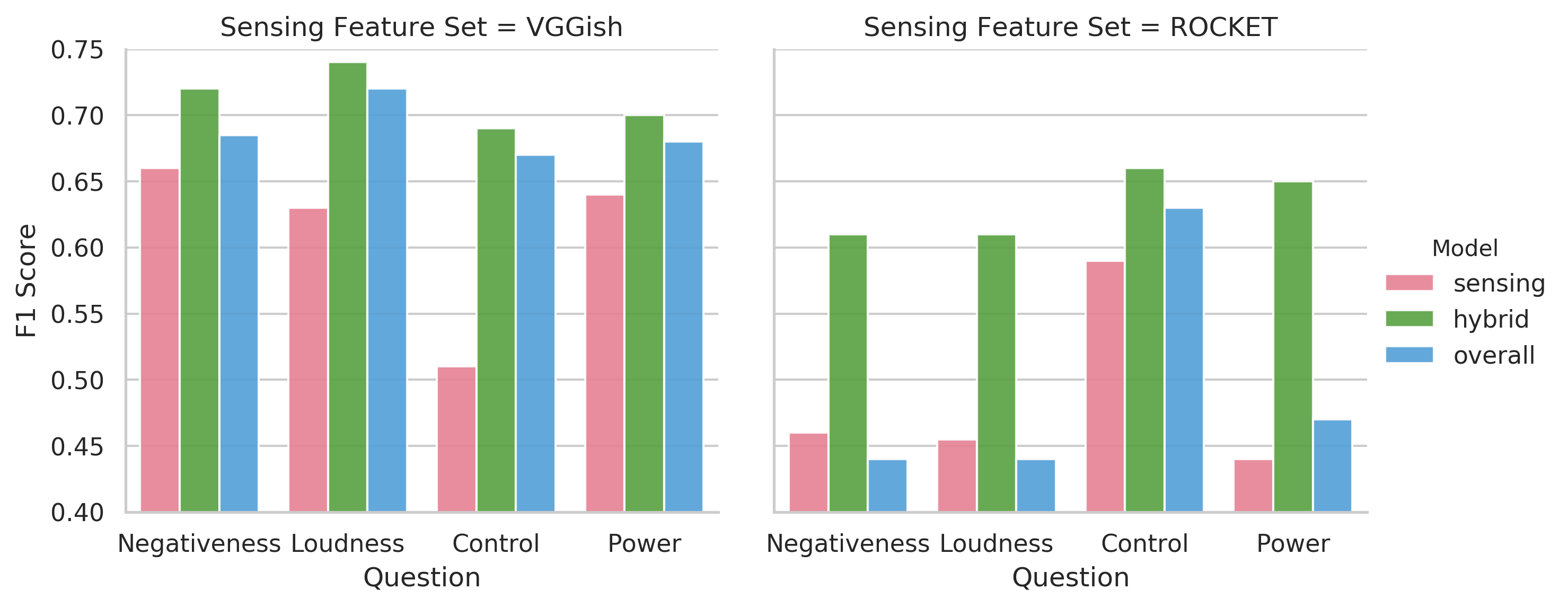}
\end{subfigure}
\caption{The top-2 F1 scores for the models that are trained using the transformed sensing features. The left figure shows the performance based on the VGGish-based transformed features. The right figure are the performances based on the features generated by ROCKET.}
\label{fig:top2_both}
\end{figure}

 We also show the top-2 F1 scores in Table~\ref{fig:top2_both}. As illustrated, the models trained on the VGGish-based sensing features outperform the other models. Also note that, we try to train the models using only the main mobile sensing data, without any further feature transformations. However, we find the prediction performance not as promising, obtaining a top-1 F1 score of 41\% at best among all the four EMA questions, even with a hybrid model. 
\section{Discussions}
\label{sec:discussion}
As mentioned in Section~\ref{sec:method}, we consider a period of 24 hours prior to each EMA to generate sensing main features and then extract a further set of higher dimensional features. The reason for considering 24 hours of data is that the first question of the EMA asks if a participant has experienced to hear voices \textit{right now}. Since it is a time-sensitive question, we believe using signals relatively closer to the answer time would provide us with more relevant and powerful signals. The challenge, however, is to generate features to improve the prediction performance. Instead of hand-crafting features, we introduce a new way of transforming the mobile time series to the time domain and banded spectrograms, and then extracting features from the VGGish model~\cite{45857, 45611}. By looking at the spectrogram generated from the transformed sensing time series in Figure~\ref{fig:audio_convert}, we learn that the differences in the values of data points in the original time series are clearly demonstrated in the banded spectrogram. Hence, the features extracted from the VGGish are reliable representations of the data in a higher-dimensional space. Further, in Section~\ref{sec:results}, we observe that these transformed features improve the prediction performance. Humans are complex systems. It is a challenging task to understand human behavior from features in low-dimensional spaces. Transformations to higher dimensions empower machine learning models better infer human behavior and therefore make better predictions. 

As reported in Section~\ref{sec:results}, the VGGish transformations of mobile sensing features outperform the ROCKET transformations. According to ROCKET authors~\cite{dempster2019rocket}, ROCKET outperforms other approaches in time-series classification tasks. The authors show that as they increase the number of randomly generated features from time series using random convolutional kernels, the performance increases. However, in our specific scenario, to have a fair comparison with the VGGish-based transformation, we generated only 128 features using 64 random kernels from ROCKET. As the features generated by ROCKET are based on random kernels, there is a possibility that if the number of computed features is not high, then the generated representations may not be powerful enough. On the other hand, hyperparameters and kernels used in the VGGish model are learned and experimentally tuned to find representational patterns in the data. Therefore, we see that the features from VGGish help the prediction model obtain a higher performance, although ROCKET performs better in terms of the computational complexity. The results mostly show that other approaches such as the VGGish transformation for time series achieve a higher classification performance if the number of transformed features is not too high. We believe this observation would help future work to broadly investigate how transforming time series features better contributes to improving the prediction performance in complex problems.

Regarding the architectures of the models described in Section~\ref{sec:prediction}, sometimes we set a high dropout rate, especially for earlier layers where the size of layers is also high (e.g., 0.6 dropout rate for the input layer of the overall model). The reason behind this is that while our training sample is relatively small, the multimodal features we extract are numerous. Therefore, we find that if we set the dropout rate to be a small number, we encounter the exploding gradient problem. We observe that a higher dropout rate, especially for earlier layers, helps the models perform better.

An implication of the research could be a deployed mobile system that can help healthcare professionals, psychologists and psychiatrists give timely treatments to AVH patients. The application could continuously monitor a patient’s behavior and periodically assess for an auditory hallucination event and its valence. The valence value may be thresholded to identify the severity of the AVH event. These periodic assessments can either be done automatically or could be driven by the patients via a mobile application. The results of the assessments then would be sent to the patient’s primary care for further evaluations if any psychosis or relapses have happened. In the case of an assessment requested by the patient, the patient can also send additional information about their AVH experiences via text, audio, or even video recordings to their care providers. The healthcare providers could benefit from all this information to better understand the situation and give effective treatments.

Finally, any form of data collected from individuals begs the question of data ethics and privacy. While it is widely shown that the usage of mobile sensing could benefit both the patients and clinicians in the field of mobile health, there is still a need for the discussion of risks related to such sensitive data, such as the raw audio diaries we use for our modeling purposes. One way of handling the data privacy issue is by limiting the access of raw data so that it does not leave the participants' device at all. For instance, in Section~\ref{sec:results}, we showed that our hybrid model which is trained on features extracted from other pre-trained models provides the best prediction performance. The latest iterations of mobile phones are powerful enough and are mostly equipped with AI chips that make it easy to run neural models on the phone. Therefore, it is possible to train a predictive neural model and run a frozen architecture (i.e., up to several certain layers) on the phone. Then, instead of uploading the entire raw data to web servers, we can first feed the raw data into the on-device pre-trained model and thereafter, upload the values extracted from the model to the server for further predictions. This way, the raw data never leaves the participants' devices.
\section{Related Work}
\label{sec:related_work}
Mobile technologies offer the advantage of passively collected context based in-situ data, which can be used to make a variety of inferences about the users' behavior. As a result, mobile sensing has the potential to be used for numerous purposes. For example, a number of studies find associations between mobile sensing and behavioral markers of mental health such as anxiety \cite{huang2016assessing,saeb2017mobile, info:doi/10.2196/10101, boukhechba2017monitoring}, depression \cite{Zakaria2019, wang2014studentlife,wang_tracking_2018, mehrotra2016towards, saeb2016relationship}, mood \cite{likamwa_moodscope:_2013, jaques2015predicting, hammer2014exploiting}, and other wellbeing and personality-related topics \cite{Wang:2018:SBC:3279953.3264951, Sandstrom2017, Mnsted2018, 10.1145/3491102.3502043,10.1145/3543194}. With regards to the subject of hallucinations, most of the prior studies use retrospective measures to study them \cite{deLeedeSmith2013}. However, this leads to the obvious loss of information due to recall effect and bias \cite{BenZeev2012, stone1999science, Schwarz1999}. Real-time monitoring is a solution to such problems and EMA can help achieve that. EMA or experience sampling based research on hallucination helps in capturing current data (as opposed to retrospective data) as people can report experience as it happens. Prior researchers report several relationships between time of day \cite{bless2017sa110}, worry \cite{hartley2014experience}, emotional state \cite{delespaul2002determinants} and hallucinations in EMA-based research. With the addition of mobile sensing, we can not only capture self-reported responses and audio diaries from the participants, but can also obtain behavioral data perhaps useful for monitoring and assessing early symptoms. Smelror et al. \cite{Smelror2019} study the feasibility and acceptability of mobile phone applications in characterizing AVH. They develop a mobile app that utilizes EMA to obtain AVH-related ground truth (such as the content, severity, intensity of the AVH) from the participants throughout several time points in a day. They find that the EMA-based responses through mobile phones are able to capture the fluctuations in several of the AVH dimensions (i.e., ground truths). Other studies also find EMA to be useful to assess AVH \cite{Kimhy2017}. 

In the context of data modeling, mobile sensing data is used mostly with hand-crafted features on traditional machine learning models. However, there is an increasing number of studies using deep learning-based models as well. Deep models allow us to capture patterns and trends in the data without having to come up with manual features. Deep neural networks are applied to various mobile sensing applications, specifically: activity recognition \cite{Krishna2018, radu2018multimodal}, mental health \cite{gjoreski2017deep,jaques2017predicting, Acikmese2019}, human mobility \cite{fan2018online} and security \cite{lu2018deepauth}. Haque et al. \cite{haque2018measuring} use a deep  Casual Convolution Neural Network (C-CNN) to measure severity of depressive symptoms based on audio, text transcripts and 3D facial scans of 142 patients. They report being able to predict participant's Patient Health Questionnaire (PHQ) scores with an error rate of about 15\%. Similarly, they obtain a specificity score of 82.6\% and sensitivity score of 83.3\% when detecting major depressive disorder on the patients. Mehrotra et al. \cite{Mehrotra2018} use an auto-encoder to automatically generate features from GPS-based mobility traces to predict binary depressive states of the participants with 91\% specificity and 77\% sensitivity.
Rezaii et al. \cite{Rezaii2019} extract linguistic markers from the speech samples of 40 participants. Using linguistic analysis, they show that they are able to predict whether the person will convert to psychosis with 93\% accuracy. Specifically, they report that ``conversion to psychosis is signaled by low semantic density and talk about voices and sounds'' \cite{Rezaii2019}. Researchers are also exploring transfer learning \cite{torrey2010transfer} to make predictions on mental health and wellbeing. Alsentzer et al. \cite{alsentzer2019publicly} share pre-trained BERT models trained on clinical texts and show that their domain specific model yields better performance in clinical Natural Language Processing (NLP) tasks. Similarly, Trotzek et al. \cite{trotzek2018utilizing} use pre-trained word vectors for early detection of depression in 752 users based on their linguistic metadata from social media. 

\section{Conclusion}
\label{sec:conclusion}
We studied N=435 individuals, who report hearing voices, and assessed auditory verbal hallucination. We primarily collected ecological momentary assessments and audio diaries about the content of the voices along with sensing data from a mobile application. We collected the multimodal data in two forms: active data that refers to the audio diary that participants record to implicitly describe and talk about their AVH experiences; and passive data that refers to data that our mobile sensing app collects in the background. We then experimented with how predictive language-related and contextual cues from the audio diary and mobile sensing data are of an auditory verbal hallucination event. Using transfer learning and data fusion techniques, we trained a hybrid neural model to predict the valence of the voice. The prediction performance of the hybrid model is 54\% and 72\% in terms of top-1 and top-2 F1 score, respectively. To improve the prediction performance, we proposed a new approach of transforming sensing data to a banded spectrogram to extract features in a higher-dimensional space using VGGish to make the prediction model have higher degrees of freedom to estimate the output with better accuracy. Our work as a proof of concept represented the first time that mobile data is used to assess auditory verbal hallucination.
\section*{acknowledgment}
\label{sec:acks}
This study was supported by a grant from the National Institute of Mental Health with the project number of R01MH112641.

\bibliographystyle{ACM-Reference-Format}
\bibliography{paper_bib}

\end{document}